# The entropy of "strange" billiards inside n–simplexes

**Thomas Schürmann**

Department of Theoretical Physics, University of Wuppertal, Germany

**Ingo Hoffmann**

Department of Chemical Engineering, University of Dortmund, Germany

**March 17, 1995**

### Abstract

In the present work we investigate a new type of billiards defined inside of $n$–simplex regions. We determine an invariant ergodic (SRB) measure of the dynamics for any dimension. In using symbolic dynamics, the (KS or metric) entropy is computed and we find that the system is chaotic for all cases $n > 2$.

## 1 Introduction

Let $Q$ be a bounded region in $\mathbb{R}^n$ with piece–wise smooth boundary ($n \geq 2$). A billiard in $Q$ is a dynamical system generated by the uniform linear motion of a material point inside $Q$ with a constant velocity, and with reflection at the boundary such that the tangential component of the velocity remains constant and the normal component changes sign[1]. The phase space of a billiard consists of all possible pairs $(\vec{q}, \vec{v})$ where $\vec{q} \in Q$ and $\vec{v}$ is the velocity vector, i.e. an element of the unit sphere in $n$-dimensional space. Let $M$ be the set of the points $x = (\vec{q}, \vec{v})$, and $F_t : M \longrightarrow M$ the flow on $M$. This dynamics is most conveniently replaced by a map $G_t : \partial M \longrightarrow \partial M$ which maps $(\vec{q}_n \in \partial Q, \vec{v}_n)$ onto $(\vec{q}_{n+1} \in \partial Q, \vec{v}_{n+1})$. Here, $\vec{q}_n$ is on the boundary of $Q$ and $\vec{v}_n$ is the velocity before the point hits the boundary, while $\vec{v}_{n+1}$ is the velocity after reflection and $\vec{q}_{n+1}$ the next point where the boundary is hit.

Several statements are proved about billiards moving in various geometries[1][2]. For example, the entropy of billiards inside $n$–dimensional polyhedrons is zero, while some other billiards are among the simplest hamiltonian systems proven to be ergodic and mixing. Thus the investigations of billiards are interesting from the dynamical system point of view. Also, a number of other problems of physics can be reduced to the study of billiards.

In the present work we investigate a new type of dynamical systems, so–called "strange" billiards, moving in $n$-simplex regions. They are similar to the billiards discussed above, but the reflection rule at the boundary is substituted by a "strange" rule defined in the following section. Although the dynamics will seem somewhat artificial, there are several relations to *hybrid dynamical systems* recently investigated in modelling chemical manufacturing[3][4].

In section 3 we determine the invariant ergodic measure (also called SRB measure) of the dynamics for any dimension. A generating partition is considered and we find that its corresponding symbolic dynamics is ergodic. Further we determine the (KS or "metric") entropy and find that it is positive for all cases $n > 2$. Finally the topological entropy is derived.

## 2 Dynamics of "strange" billiards

As mentioned in the introduction, "strange" billiards are defined as special standard billiards where the reflection rule is substituted by a "strange" rule. For the definition, let $S_n$ be the standard $n$–simplex embedded in $\mathbb{R}^n$

$$S_n = \left\{ \vec{x} \in \mathbb{R}^n \middle| \sum_{i=1}^n x_i = 1, \ x_j \geq 0, \quad \text{for} \quad j = 1, \ldots, n \right\}. \tag{1}$$

Further we denote by $S_{k,n-1}$ the $k$–th face of the boundary of $S_n$. They are $(n-1)$–simplexes

$$S_{k,n-1} = \left\{ \vec{x} \in \mathbb{R}^n \middle| \ x_k = 0, \ \vec{x} \in S_n \right\}, \qquad \text{for} \quad k = 1, \ldots, n. \tag{2}$$

Note that the boundary of $S_n$ is just the union $\partial S_n = \bigcup_{i=1}^n S_{i,n-1}$ of the faces.

As long as $\vec{x}$ is not on the boundary of $S_n$, the dynamics is as in the ordinary billiard: $(\vec{x}, \vec{v}) \xrightarrow{F_t} (\vec{x} + \vec{v}t, \vec{v})$. But in contrast the reflection at the boundary is not specular. Instead, for each face $S_{k,n-1}$ there is a fixed velocity $\vec{v}_k$ under which the trajectory leaves the face, independently of the direction of the incident velocity. Furthermore, we assume that these velocities are not independent, in particular $\vec{v}_k = \vec{e}_k - \vec{\rho}$ where $\vec{e}_k$ is the $k$–th canonical unit vector in $\mathbb{R}^n$ and $\vec{\rho}$ an element of the interior of $S_n$. Note that the dynamics is completely defined by the choice of $\vec{\rho}$.

A realization of this model is a set of $n$ containers served by a single server. The content $x_i$ in the $i$–th container decreases by a rate $\rho_i$, while the server can refill with unit rate. To start with, $\sum_{i=1}^n x_i = 1$ and the server is in position $i_0$. It remains there filling container $i_0$ until another container becomes empty. At this time the server switches instantaneously and remains in the new position until the next container is emptied, and so on[3].

For the definition of a symbolic dynamics we first consider "Poincaré" sections by restriction of the trajectory to the boundary of $S_n$. Therefore let us denote by $t_m$ the time where the trajectory $\vec{x}(t)$ strikes the boundary $\partial S_n$ at the $m$–th times. The time interval $\Delta t_m \equiv t_{m+1} - t_m$ between two successive hits of the boundary at $\vec{x}(t_m)$ and $\vec{x}(t_{m+1})$ is determined unique from $\vec{x}(t_m)$. Precisely, suppose that the trajectory hits at time $t_m$ the $k$–th face i.e. $\vec{x}(t_m) \in S_{k,n-1}$, then $\Delta t_m$ is determined by simple geometrical arguments and is

$$\Delta t_m = \min_{i \neq k} \left( \frac{x_i(t_m)}{\rho_i} \right). \tag{3}$$

Then we formally write the Poincaré map by

$$\vec{x}(t_{m+1}) \equiv G(\vec{x}(t_m)) = \vec{x}(t_m) + (\vec{e}_k - \vec{\rho}) \Delta t_m, \tag{4}$$

where $\vec{e}_k$ is corresponding to the face $S_{k,n-1}$ stroken at time $t_m$.

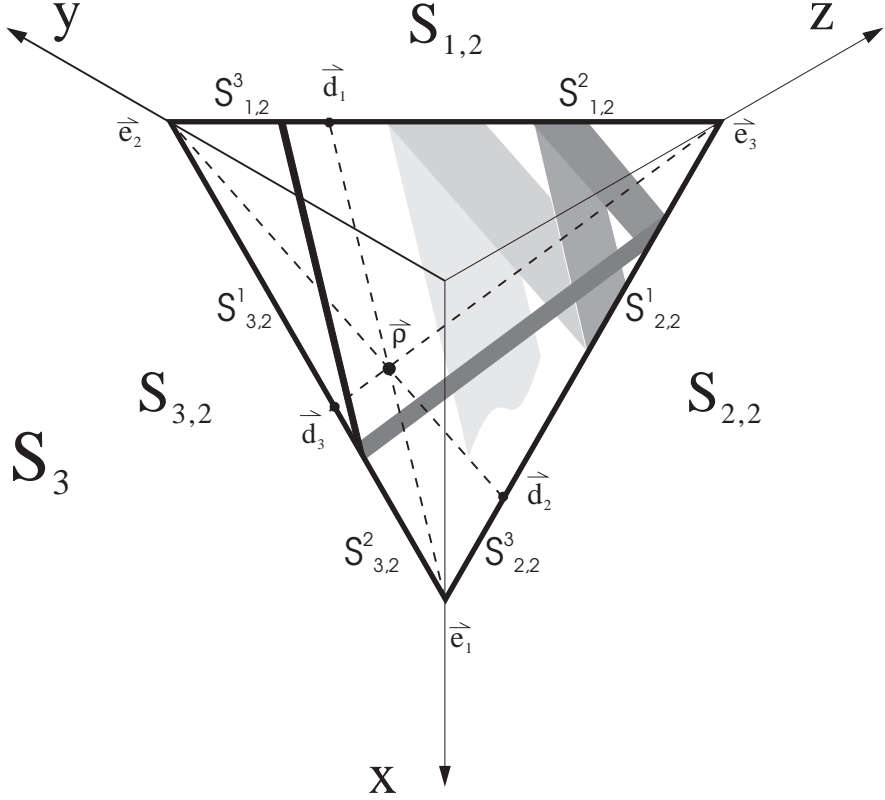

Figure 1: Illustration of the state space of a "strange" billiard inside a 3–simplex. The spreading of nearby trajectories is presented by the grey area. The transition point $\vec{\rho}$ parameterizes the dynamics $G$.

Of special interest are the points $\vec{d}_k \in S_{k,n-1}$ from which the trajectory hits the opposite corner. It is defined by[1]

$$\vec{e}_k = \vec{d}_k + (\vec{e}_k - \vec{\rho}) \alpha, \qquad \alpha \in \mathbb{R}, \qquad \text{and} \tag{5}$$
$$\langle \vec{d}_k, \vec{e}_k \rangle = 0. \tag{6}$$

---
[1] $\langle , \rangle$ denotes the canonical scalar product in $\mathbb{R}^n$

This gives $(\rho_k - 1)\alpha + 1 = 0$, and thus

$$\alpha = \frac{1}{1 - \rho_k}, \qquad \text{i.e.} \tag{7}$$

$$\vec{d}_k = \frac{1}{1 - \rho_k} (\rho_1, \rho_2, \ldots, 0_k, \ldots, \rho_n). \tag{8}$$

We have thus proven the following lemma:

**Lemma 2.1** *Let $\vec{d}_k \in S_{k,n-1}$ be defined by eq.(8). It follows that*

$$\vec{d}_k \xmapsto{G} \vec{e}_k, \tag{9}$$

*and the time step associated with this map is $\Delta t = \alpha$, where $\alpha$ is given by eq.(7).*

We note that the trajectories starting with a point $\vec{x} \in S_{k,n-1}$, see eq.(4) are always parallel to the line between $\vec{d}_k$ and $\vec{e}_k$ through the point $\vec{\rho}$, because of the same directional vector of the trajectory.

Finally let us define a symbolic dynamics. Each time $t_m$ when a face $S_{\omega_m,n-1}$ is hit by the particle we write down the index $\omega_m$ of the actual face. Then any point $\vec{x}$ which gives rise to a trajectory with $\vec{x}(t_0) = \vec{x}$ and $\vec{x}(t_m) \in S_{\omega_m,n-1}$ is coded by an infinite sequence

$$\omega = \{\omega_0, \omega_1, \ldots\}, \qquad \omega_k \in \{1, \ldots, n\}. \tag{10}$$

The dynamics on the space of sequences is indeed defined by the the ordinary shift transformation $\omega' = \sigma(\omega)$, where $\omega'_k = \omega_{k+1}$. For the computation of the entropy of $G$ one can equivalently determine the entropy of the shift $\sigma$ if the partition is generating[6].

In the following section we first determine an invariant ergodic measure of the map $G$. Further we show that the partition beyond the transformation $\sigma$ is generating and that $\sigma$ represents an ergodic Markov shift. The entropy of the map $G$ is then straightforward determined by computing the entropy corresponding to the Markov measure.

## 3 Determining invariant measures and entropies

For what follows we give here a short repetition of the theoretical framework about measures, probabilities and ergodicity. We consider the invariant set $\partial S_n$. For each measurable subset $B$ of $\partial S_n$, the symbol $\Lambda(B) = \int_B d\Lambda(\vec{x})$ denotes the Lebesgue measure of $B$. We look for an invariant probability measure which is absolutely continuous with respect to $\Lambda$, and for its density $\mu$. In nonlinear dynamics a $\Lambda$-density $\mu$ is called an invariant probability density function w.r.t. the dynamics of the system. In view of the natural given partition $\mathcal{P} = \{S_{i,n-1} | i = 1, \ldots, n\}$ of $\partial S_n$ we consider the special subsets $S_{i,n-1}$. They are disjoint, apart from their edges which have measure zero. The measure of subsimplex $i$ can be expressed as

$$\lambda(S_{i,n-1}) = \int_{S_{i,n-1}} \mu(\vec{x}) \, d\Lambda(\vec{x}) \tag{11}$$

The probability density function $\mu(\vec{x})$ is called to be invariant w.r.t. the map G, if

$$\lambda\left(G^{-1}(B)\right) = \lambda(B) \tag{12}$$

for all $B \subseteq \partial S_n$. Suppose that the invariant probability density function $\mu$ cannot be written as the superposition $\mu = \alpha\mu_1 + (1-\alpha)\mu_2$ with $0 < \alpha < 1$, $\mu_1 \neq \mu_2$. Then $\mu$ is called indecomposable or ergodic. Regardless of the initial probability density function, provided it was absolute continuous w.r.t. Lebesgue, the system will asymptotically be described by the invariant probability density function $\mu^*$.

Now let us construct the Frobenius-Perron operator and determine the invariant probability density function $\mu^*$ on $\partial S_n$. By $\mathbf{1}_A(x)$ we denote the characteristic (or indicator) function for a set $A$ simply defined by

$$\mathbf{1}_A(x) = \begin{cases} 1 & \text{if } x \in A \\ 0 & \text{if } x \notin A \end{cases}. \tag{13}$$

Then we consider the ansatz of a piece–wise constant density

$$\mu(\vec{x}) = \sum_{i=1}^{n} \mu_{i,n-1} \, \mathbf{1}_{S_{i,n-1}}(\vec{x}), \qquad \mu_{i,n-1} \in \mathbb{R}^+. \tag{14}$$

For the particular sets $S_{k,n-1}$, $k = 1, \ldots, n$, we find by the above definition

$$\lambda(S_{k,n-1}) = \mu_{k,n-1} \Lambda(S_{k,n-1}) \tag{15}$$

and

$$\lambda(S^l_{k,n-1}) = \mu_{k,n-1} \Lambda(S^l_{k,n-1}). \tag{16}$$

Here, $\Lambda(A)$ denotes the Lebesgue measure of set $A$ in dimension $(n-2)$ and $S^l_{k,n-1}$ is the subset of $S_{k,n-1}$ which is mapped under $G$ to $S_{l,n-1}$, see fig.(1). Note that $G^{-1}$ is not a single valued map. Invariance of the measure $\lambda$ requires then a fixed point equation for the densities:

$$\mu^*_{l,n-1} = \sum_{k=1}^{n} \frac{\Lambda(S^l_{k,n-1})}{\Lambda(S_{k,n-1})} \, \mu^*_{k,n-1}, \qquad l = 1, \ldots, n. \tag{17}$$

**Lemma 3.1** *In the above constructed state space and under the mentioned assumptions it is*

$$\frac{\Lambda(S^l_{k,n-1})}{\Lambda(S_{k,n-1})} = \langle \vec{d}_k, \vec{e}_l \rangle. \tag{18}$$

**Proof:** The simplexes $S_{k,n-1}$ and $S^l_{k,n-1}$ have in common an entire $(n-3)$-dimensional subsimplex, namely the face opposing the corner $\vec{d}_k$ of $S^l_{k,n-1}$. We call this face $S_{k;l,n-2}$. Therefore we have

$$\frac{\Lambda(S^l_{k,n-1})}{\Lambda(S_{k,n-1})} = \frac{\text{dist}\left(\vec{d}_k, S_{k;l,n-2}\right)}{\text{dist}\left(\vec{e}_l, S_{k;l,n-2}\right)}, \tag{19}$$

where

$$\text{dist}(\vec{x}, A) = \min\{|\vec{x} - \vec{y}| \,:\, \vec{y} \in A\}. \tag{20}$$

The vectors $\vec{d}_k - \vec{y}_1$ and $\vec{e}_l - \vec{y}_2$ (where $\vec{y}_1$ and $\vec{y}_2$ minimize the denumerator and denominator in eq.(19), respectively) are both in the linear space spanned by $S_{k,n-1}$ and perpendicular to $S_{k;l,n-2}$. They are thus parallel vectors, and the ratio of their lengths is equal to the ratio of any one of their components. Taking in particular the $l$-th component (for which $(\vec{y}_1)_l = (\vec{y}_2)_l = 0$ and $(\vec{e}_l)_l = 1$) we find eq.(18).

With lemma 3.1 and eq.(17) we get immediately the following theorem:

**Theorem 3.1** *Any measure $\lambda$ which has constant density $\mu_{k,n-1}$ on each subsimplex $S_{k,n-1}$ evolves under the application of the map $G$ according to*

$$\mu_{l,n-1} = \sum_{k=1}^{n} (\vec{d}_k)_l \; \mu_{k,n-1}. \tag{21}$$

Note that by using $\sum_{i=1}^{n} \rho_i = 1$ and the fact that $\vec{\rho}$ is in the interior of $S_n$, one easily verifies the identity $\sum_{l=1}^{n}(\vec{d}_k)_l = 1$ for $l = 1, \ldots, n$, proving that the matrix $p_{lk} = (\vec{d}_k)_l$ is indeed a stochastic matrix.

The following theorem yields a fixed point of eq.(21)

**Theorem 3.2** *The piece–wise constant measure*

$$\lambda^*(S_{i,n-1}) = \frac{1}{d} \, \rho_i(1-\rho_i), \qquad i = 1, \ldots, n, \tag{22}$$

$$d = \sum_{i=1}^{n} \rho_i(1-\rho_i), \tag{23}$$

*is a properly normalized probability measure and invariant under $G$.*

**Proof:** By straightforward computation.

For $n = 3$, this has been derived already in [3]. Indeed the above measure is absolutely continuous and also unique due to the Perron–Frobenius theorem[7]. Such measures are called SRB measures. Its corresponding density is given by the relations (15,16). Finally, we state our main result:

**Theorem 3.3** *Let $G : \partial S_n \to \partial S_n$ be defined by equation (4). Let also the dynamics be determined by the point $\vec{\rho}$. Then the entropy $h_\mu$ corresponding to the map $G$ is*

$$h_\mu = \frac{1}{d} \sum_{i=1}^{n} \rho_i (1 - \rho_i) \, \log \frac{1 - \rho_i}{\rho_i} \qquad \text{with} \qquad (24)$$

$$d = \sum_{i=1}^{n} \rho_i (1 - \rho_i).$$

**Proof:** Let us consider the finite partition $\mathcal{P} = \{S_{i,n-1} | \, i = 1, \ldots, n\}$ of $\partial S_n$. Since the boundaries $\partial S_{i,n-1}$ of its elements are invariant w.r.t. the map $G$, i.e. the boundaries of the subsets $S_{i,n-1}$ can not be mapped in interiors, the partition $\mathcal{P}$ is a *Markov partition*[5]. Markov partitions for which all intersections $S_{i,n-1} \cap G(S_{j,n-1})$ and $S_{i,n-1} \cap G^{-1}(S_{j,n-1})$ are connected are generating partitions[6]. This is indeed the case for the above system. Also our partition has the generating property. The symbolic dynamics $\sigma$ (c.f. sec.2) is thus equivalent to a discrete time Markov process, i.e. the probability of hitting face $S_{i,n-1}$ at time $t_m$ is only dependent on the preceding face hit at $t_{m-1}$. For the complete statistical characterization of the system, let $p_i = \rho_i (1 - \rho_i)/d$, $i = 1, \ldots, n$, be the weight of the $i$-th subsimplex. The (conditional) transition probabilities for transitions $j \to i$ are given by the matrix $p_{ij} = (\vec{d}_j)_i$. The shift $\sigma$ on the space of sequences is a measure preserving transformation for the Markov measure defined by $p_{ij}$ and initial vector $p_j$, satisfying $p_i = \sum_j p_{ij} \, p_j$. Also the transformation $\sigma$ is ergodic because the chain is irreducible, i.e. for all pairs of states $i, j$, there exists a $m > 0$ with $p_{ij}^{(m)} > 0$ (e.g. $m = 2$). Since the entropy of an ergodic Markov shift is[6]

$$h_\mu = - \sum_{i,j} p_j \, p_{ij} \, \log p_{ij}, \qquad (25)$$

by inserting the above measures and doing some algebraic simplifications we get equation (24).

Hence for $n > 2$ we get positive entropy. Since all transitions $i \to j$ with $i \neq j$ are allowed, while $i \to i$ is forbidden, the topological entropy is $h_{top} = \log(n - 1)$.

## 4 Summary and outlook

Inspired by hybrid dynamical systems recently investigated in modelling chemical manufacturing, we introduced dynamical systems on $n$–simplex geometries, generalizing thereby the results obtained in [3] to arbitrary $n$.
By use of generating partitions it was possible to define symbolic Markov dynamics. Due to the ergodicity of the corresponding Markov shift $\sigma$ it was possible to compute the metric entropy with help of the underlying SRB measure.
In the considered system the velocities $\vec{v}_k = \vec{e}_k - \vec{\rho}$ of the particle are not independent, but $\vec{v}_k - \vec{v}_l = \vec{e}_k - \vec{e}_l$. One might ask what happens in the more general case where the velocities still depend only on the index $k$ of the hit boundary simplex, but where they are not restricted by this constraint. By the Perron–Frobenius theorem there also exists an unique probability measure but its algebraic expression is not as simple as in the case above, especially for larger values of $n$.

A further class of dynamical systems is obtained by choosing the velocities not only dependent on the faces at the actual "reflection" as in the present work, but also depending on the particular position where the trajectory hits the boundary[8].

In such systems one also observes simplex–like invariant sets with singularities but their dynamical properties are quite different. In [8] we will introduce briefly the generalized dynamical systems and will also discuss possible applications to chemical, bio–technological manufacturing and socio–economical systems.

## Acknowledgment

The interest of Markov partitions to the dynamical system is due to P. Grassberger. We like to thank him for helpful and clarifying discussions. Also we want to thank E. Grycko for interesting discussions.

## References

[1] Ya. G. Sinai, *Dynamical systems with elastical reflections*, Uspeki. Math. Nauk **27** (1972) 137

[2] L. A. Bunimovich, Ya. G. Sinai and N. I. Chernov *Statistical properties of two–dimensional hyperbolic billiards*, Russ. Math. Surv. **46:4** (1991) 47

[3] C. Chase, J. Serrano and P. J. Ramadge, *Periodicity and chaos from switched flow systems: contrasting examples of discretely controlled continuous systems*, IEEE Trans. Automat. Contr. **38** (1993), 70

[4] P. J. Ramadge, *On periodicity of symbolic observations of piece–wise smooth discrete-time systems*, IEEE Trans. Automat. Contr. **35** (1990), 807-813

[5] D. H. Mayer, *Approach to equilibrium for locally expanding maps in $\mathbb{R}^n$*, Commun. Math. Phys. **95** (1984) 1

[6] Ya. G. Sinai, TOPICS IN ERGODIC THEORY, Princeton University Press, 1994

[7] R. Mañé, ERGODIC THEORY AND DIFFERENTIABLE DYNAMICS, Springer, 1987

[8] I. Hoffmann, T. Schürmann, *The entropy of hybrid dynamical systems*, in preparation